\documentclass[aps,pre,twocolumn,showpacs]{revtex4}

\usepackage{graphicx}
\usepackage{amssymb}
\usepackage{amsmath}

\newcommand{\ie}{i.e.}
\newcommand{\Ttt}{{\mathcal T}}

\begin{document}

\date{\today}
\title{Cluster persistence in one-dimensional diffusion--limited
cluster--cluster aggregation}  
\author{E. K. O. Hell\'en}
\email{ehe@fyslab.hut.fi}
\author{P. E. Salmi} 
\email{psa@fyslab.hut.fi}
\author{M. J. Alava}
\email{mja@fyslab.hut.fi}
\affiliation{Laboratory of Physics, Helsinki University of Technology,
P. O. Box 1100, FIN-02150 HUT, Finland}
\date{\today}

\begin{abstract}
The persistence probability, $P_C(t)$, of a cluster to remain
unaggregated is studied in  cluster-cluster
aggregation, when the diffusion coefficient of a cluster depends on
its size $s$ as $D(s) \sim s^\gamma$. In the mean-field the 
problem maps to the survival of three annihilating random walkers with
time-dependent noise correlations. For $\gamma \ge 0$ the motion of 
persistent clusters becomes asymptotically irrelevant and the
mean-field theory provides a correct description.
For $\gamma < 0$ the spatial fluctuations remain relevant and the
persistence probability is overestimated by the random walk theory. 
The decay of persistence determines the small size tail of the cluster
size distribution. For $0 < \gamma < 2$ the distribution
is flat and, surprisingly, independent of $\gamma$.

\end{abstract}



\maketitle

\section{Introduction} \label{intro}

Aggregation models are useful in describing various
phenomena from chemical engineering, material sciences,
atmosphere research to even
astrophysics~\cite{Meakin:PhysicaScripta46,Friedlander_book,Family_Landau}.
One general property of these models is that  
they lead to dynamic scale-invariance: when all the lengths are scaled
by the characteristic length, the system looks the same at
different times.  
Lately, studies of first passage problems~\cite{rednerinkirja} under
the name
persistence~\cite{Majumdar:CurrSciIndia,Majumdar:PRL77,Derrida:PRL77}
have shown that not 
necessarily all the properties of a dynamically scaling system 
are characterized by a single scale~\cite{Bray:PRE62}. Here we address
the probability of a cluster to remain intact
in an aggregation system and show how this quantity and the associated
length scale relate to the physically relevant issue of the shape of
the cluster size distribution.  

In an aggregation system one can define many first-passage problems
and related quantities~\cite{omapers}. 
We study the probability that a cluster
has not aggregated with any other one before time
$t$~\cite{Hellen:EPL}. This probability is called cluster persistence
and denoted by $P_C(t)$. Similar problems considering uninfected
walkers in one-dimensional reaction-diffusion
systems~\cite{ODonoghue:CM1} and Potts model~\cite{ODonoghue:CM2} have
recently been shown to display interesting behavior. We concentrate on 
diffusion--limited cluster--cluster aggregation (DLCA) in one
dimension, where the dynamics is dominated by spatial
fluctuations~\cite{Kang:PRA30}. For high dimensional 
systems these may be neglected, and on the mean-field level, valid for
dimensions higher than the upper critical dimension, aggregation is
well understood~\cite{vanDongen:PRL54,vanDongen:JSP50,vanDongen:PRL63}. 

The DLCA model is defined so that the nearest neighbor occupied 
sites in a lattice are identified as a cluster. Each cluster diffuses
with a size dependent diffusion constant, $D(s) \sim s^\gamma$, where
$\gamma$ is the diffusion exponent. If a cluster collides with another
one, the two clusters are irreversibly merged together and the
aggregate diffuses either faster ($\gamma>0$) or slower ($\gamma<0$)
than  either of the colliding clusters. 
While the size independent diffusion ($\gamma=0$) is exactly solvable
in one dimension, it forms a marginal case between two completely
different aggregation mechanisms~\cite{Hellen:PRE62}. 
We study here the more physically interesting problem with $\gamma
\neq 0$.

The aim is to study the dependence of cluster persistence
on the diffusion exponent $\gamma$ and extend the study presented in
Ref.~\cite{Hellen:EPL}. We also pay attention to the random-walk (RW)
problems that ensue as on a mean-field level the problem is reduced 
to the survival of three annihilating random walkers.
While the $\gamma=0$ case is readily solvable
by various methods \cite{Fisher:JSP34,Derrida:PRE54}, already the case
of three annihilating particles with unequal diffusion constants
is rather involved~\cite{Fisher:JSP53}. Here $\gamma \neq 0$ leads to
time dependent diffusion coefficients, and 
we derive a Fokker-Planck~(FP) equation for the survival of these
particles. For $\gamma \ge 0$ its analysis yields an algebraically decaying
survival probability, $P_{\rm surv}(t) \sim t^{-\theta_{\rm
RW}(\gamma)}$. The survival exponent $\theta_{\rm RW}$ is   
discontinuous and non-monotonic as it is given by $\theta_{\rm RW}(\gamma) =
2/(2-\gamma)$ for $0 < \gamma < 2$ and $\theta_{\rm RW}(0) = 3/2$.
The numerical comparison of the survival and persistence probabilities
validates the theory and hence $P_C(t) \sim t^{-\theta_C}$ with
$\theta_C = \theta_{\rm RW}$. 

For $\gamma < 0$ simulations show that both the survival and
persistence probabilities decay stretched exponentially as $\exp (-C
t^{\beta})$. The Fokker-Planck equation is not amenable to analytic 
analysis, so we use a Lifshitz tail argument to understand the survival. 
Such heuristic arguments and numerics suggest
a stretching exponent $\beta_{\rm
RW}(\gamma) = -\gamma/(4-2\gamma)$. 
The Lifshitz tail argument indicates that the exponent is affected by
the fluctuations in the motion of the particles that neighbor the
surviving one. These are taken into account only approximately in the
mean-field theory and for the DLCA numerics gives $\beta_C =
-2\gamma/(6-3\gamma)$.  A closer examination reveals that also the
distance distribution between the particles surrounding a surviving
one in the mean-field model scales in a different way than the
corresponding distribution of the DLCA.

In addition, we show how the cluster persistence
is related to the cluster size distribution.
To clarify the connection, consider the dynamic scaling
in DLCA. Both simulations and experiments show that the cluster size
distribution $n_s(t)$ (the number of cluster 
of size $s$ per lattice site at time $t$) scales 
as~\cite{Meakin:PhysicaScripta46} 
\begin{equation}
n_s(t) = S(t)^{-2} f\left(\frac{s}{S(t)}\right), \label{nstscaleq}
\end{equation}
where $S(t) \sim t^z$ is the average cluster size and the scaling
limit, $s \to \infty$ and $S(t) \to  \infty$ with $s/S(t)$ fixed, is
taken. In one dimension the dynamic exponent 
$z = 1/(2-\gamma)$~\cite{Miyazima:PRA36,Kang:PRA33}. For $\gamma
\ge 0$ the cluster size 
distribution is broad in the sense that the scaling function
behaves as $f(x) \sim x^{-\tau}$ as $x \equiv s/S(t) \to 0$. 
For $\gamma < 0$ the scaling function is bell-shaped and 
$f(x) \sim \exp (-A x^{-|\mu|})$ for $x \to 0$, where $A$ is a
constant. To determine the polydispersity exponent, $\tau$, which
characterizes the number of small clusters, is non-trivial 
even on a mean-field level~\cite{vanDongen:PRL54,Cueille:PRE55}
whereas the similar exponent $\mu$ 
readily follows from scaling analysis~\cite{vanDongen:JSP50}. All the
exponents $z$, $\tau$, and $\mu$ are universal, 
\ie, they do not depend on the fine details of the model. They can,
and it is natural to expect that they do, depend on the 
diffusion exponent $\gamma$.

One of the main results of this paper is that the exponents describing
the decay of the cluster persistence are related to these
universal exponents as  
\begin{subequations}
\begin{eqnarray}
\theta_C &=& (2-\tau)z 
\label{scalrelsforcsd1}\\
\beta_C  &=& |\mu|z. 
\label{scalrelsforcsd2}
\end{eqnarray}
\end{subequations}
Quite unexpectedly, the polydispersity exponent is a constant, $\tau
=0$, for $0< \gamma < 2$, but discontinuous since 
$\tau(\gamma=0) = -1$. 
The reasoning leading to the relations~\eqref{scalrelsforcsd1}
and \eqref{scalrelsforcsd2} is
universally applicable, so that the behavior of the tail of cluster size
distribution might be tackled through cluster persistence  
in other models, too.

The outline of the paper is as follows. In section~\ref{MF} the
mean-field random walk theory is formulated and the
associated Fokker-Planck equation is derived. Section~\ref{comparison}
starts by describing the simulation methods. Thereafter the mean-field
theory is validated for $\gamma \ge 0$ by comparing the survival
probability obtained from the analysis of the Fokker-Planck equation
to the simulation results of both the random walk system and the DLCA
one. For $\gamma < 0$ a similar comparison shows the effect of spatial
fluctuations, and the stretched exponential decay of the survival
probability is explained using a Lifshitz tail
argument. Section~\ref{poly} concentrates on the relation
between the persistence and the small size tail of the
cluster size distribution. The paper ends with conclusions in
section~\ref{conclusion}.

\section{Mean-Field: Reduction to a Three Particle Problem} \label{MF} 

The two clusters surrounding a persistent one will grow when
they collide with other clusters (but not with the persistent one).
The cluster in the middle will be
persistent until it collides with one of the neighbors. After this
the two remaining clusters would contribute to 
persistence only by increasing the mass of the clusters surrounding
another persistent cluster. This is negligible at
late times, since the persistent clusters will be separated by many
non-persistent ones, \ie\ $t^{\theta_C} \gg t^z$. 
In other words, the correlations in the system
grow only as $t^z$ and each persistent cluster is asymptotically
independent. Thus it is sufficient to consider only one persistent cluster
and its two neighbors. 

The collisions of the surrounding clusters make them
bigger and increase or decrease the diffusivity. We make the mean-field 
approximation that each cluster neighboring a persistent
one will grow as an average cluster does. Hence, we replace the true
process, where the surrounding clusters collide at some discrete times
$t_i$, by a continuous one, where the surrounding clusters grow as
$S(t)$. As $D(s) \sim s^\gamma$ these clusters will diffuse with
time-dependent diffusion coefficients. In the
following analysis we will ignore the possible early time crossover
effects in the growth of the average cluster size and the diffusion
coefficients of the clusters surrounding a persistent one are
taken to follow a true power-law at all times. This
will only affect the early time behavior.

The finite extent of clusters is irrelevant for cluster
persistence and we will consider
the three clusters as point-like particles from now on. 
Let $x_i(t)$ ($i=1,2,3$) denote the positions of the particles
at time $t$ such that $x_1(0) < x_2(0) < x_3(0)$. 
The motion of these is described by the
Langevin equations
\begin{equation}
\dot{x}_i(t) =  \xi_i(t), \label{Lang1}
\end{equation}
with Gaussian white noises $\langle \xi_i(t) \rangle
= 0$ and $\langle \xi_i(t) \xi_j(t') \rangle =  2 {\cal D}_i(t)
\delta_{ij} \delta(t-t')$. The overdot denotes derivative
with respect to time and the brackets an ensemble average over
different realizations.
The diffusion coefficients of the particles read as ${\cal D}_1(t)
= {\cal D}_3(t) = D_1 t^{\gamma z}$ and ${\cal D}_2(t) = D_2$. 
The meaning of a 
time-dependent diffusion coefficient, say ${\cal D}_1(t)$, is simply
that the particle $1$ will follow a simple diffusive motion with a
diffusion constant $D_1$ in the time scale 
\begin{equation}
T_1(t) =
\int_0^t {\rm d}t' {\cal D}_1(t')/D_1 = t^{\gamma z + 1}/(\gamma z
+ 1).  
\label{T1timescal}
\end{equation}

As we are interested in the survival of the middle particle ($x_2$),
the process terminates when either $x_1=x_2$ 
or $x_2 = x_3$. It is convenient to consider
the distances between the particles: $x_{12}(t) = x_2(t) - x_1(t) \ge
0$ and $x_{23}(t) = x_3(t) - x_2(t) \ge 0$. These obey similar
Langevin equations 
\begin{equation}
\label{Langforx12x23}
\begin{cases}
 {\dot x}_{12}(t) = \Gamma_{12}(t) \\
 {\dot x}_{23}(t) = \Gamma_{23}(t),
\end{cases}
\end{equation}
where $\langle \Gamma_{12}(t) \rangle = \langle \Gamma_{23}(t) \rangle
= 0$ and $\langle \Gamma_{12}(t) \Gamma_{12}(t') \rangle = 
\langle \Gamma_{23}(t) \Gamma_{23}(t') \rangle = 
2 (D_2 + D_1 t^{\gamma z}) \delta(t-t')$.
The two noises are correlated as the motion of the middle particle
affects both distances: $\langle \Gamma_{12}(t) 
\Gamma_{23}(t') \rangle = - \langle \xi_2(t) \xi_2(t') \rangle = - 2
D_2 \delta(t-t') \neq 0$. For $\gamma > 0$ the noise correlations
become asymptotically irrelevant, which is not the case for $\gamma <0$. 

To proceed, we transform Eqs.~\eqref{Langforx12x23} to a 
Fokker-Planck equation for the probability density
$\rho(x_{12},x_{23};t)$ of the two distances at time $t$. 
Due to the mutual correlations this is easiest to do by
computing the drift and diffusion coefficients 
from their definitions 
\begin{eqnarray}
&D_{i} &=  \lim_{\Delta t \to 0} \frac{1}{\Delta t}\left\langle x_{i}(t 
+ \Delta t) -x_{i}(t) \right\rangle \nonumber \\
&D_{ij} &= \frac{1}{2} \lim_{\Delta t \to 0} \frac{1}{\Delta t}\langle
[x_{i}(t + \Delta t) -x_{i}(t)][x_{j}(t + \Delta t) -x_{j}(t)]
\rangle  \nonumber
\label{diffusion}
\end{eqnarray}
and insert these to the general Fokker-Planck
equation~\cite{Risken:book} 
\begin{equation}
{{\partial \rho}\over{\partial t}} = -\sum_{i=1}^{2} {{\partial
}\over{\partial x_{i}}}  D_{i} \rho + \sum_{i,j=1}^{2} {{\partial^2
}\over{\partial x_{i} \partial x_{j} }} 
D_{ij} \rho.
\label{fokkerplanck}
\end{equation}
A straightforward calculation gives 
\begin{equation}
\frac{\partial{\rho}}{\partial t} = (D_2+D_1t^{\gamma z})
\left( 
 \frac{\partial^2{\rho}}{\partial x_{12}^2} 
 + \frac{\partial^2{\rho}}{\partial x_{23}^2}
\right)
- 2 D_2 \frac{\partial^2{\rho}}{\partial x_{12} \partial x_{23}}.
\label{FPeq} 
\end{equation}
The initial condition is now 
$\rho(x_{12},x_{23};0) = \delta(x_{12}-x_{12}^0)
\delta(x_{23}-x_{23}^0)$, where $x_{12}^0 = x_{12}(0)$ and $x_{23}^0 =
x_{23}(0)$ are the initial distances between particles. 
The termination of the process when two particles collide gives
absorbing boundary conditions along the axis, \ie,
$\rho(x_{12},0;t)=0$ and $\rho(0,x_{23};t) = 0$
for all times $t$.  

Thus the original many body problem has been reduced to the survival
of three annihilating random walkers. 
Given that one can solve Eq.~\eqref{FPeq} with the appropriate
boundary conditions, the survival probability of the middle particle
(which corresponds to the persistent cluster) can be obtained as 
\begin{equation}
P_{\rm surv}(t) = \int_0^\infty {\rm d}x_{12} \int_0^\infty 
{\rm d}x_{23}\ \rho(x_{12},x_{23};t). 
\end{equation}
When the survival probability decays algebraically, $P_{\rm
surv}(t) \sim t^{-\theta_{\rm RW}}$, the associated exponent
$\theta_{\rm RW}$ is called the survival exponent.

\section{Comparison of the Simulations and Theory} \label{comparison}

\subsection{Details of Simulations} \label{algorithms}

The DLCA simulations are done on a lattice of size $L$ with periodic
boundary conditions. Concentration $\phi$ of sites is filled with
particles and nearest neighbor particles belong to the same
cluster. The initial distribution is either
monodisperse, $n_s(0) = \delta_{1,s}$, with equal distances, $l_0$,
between neighboring clusters or random, in which case each site is
independently filled with probability $\phi$. The persistence
exponent is independent of the initial distribution but
the early time behavior of the persistence probability
depends on it~\cite{omapers}.

In the dynamical evolution a cluster 
is selected randomly and time is increased by $1/[N(t) D_{\rm
max}(t)]$. Here $N(t)$ denotes the number of clusters and $D_{\rm max}(t)$  
is the maximum of the diffusion coefficients of all the clusters at
time $t$. The cluster is moved one lattice 
spacing with cluster size 
dependent probability $D(s)/D_{\rm max}(t)$. 
If the cluster collides with another one, 
the clusters are irreversibly
aggregated together and the values of $N$ and $D_{\rm max}$ are
updated. Then a new cluster is selected and the above procedure
is repeated.

\begin{figure*}
        \centering
        \begin{minipage}[b]{.48\linewidth}
                \centerline{\hbox{
\includegraphics[angle=-90,width=\linewidth]{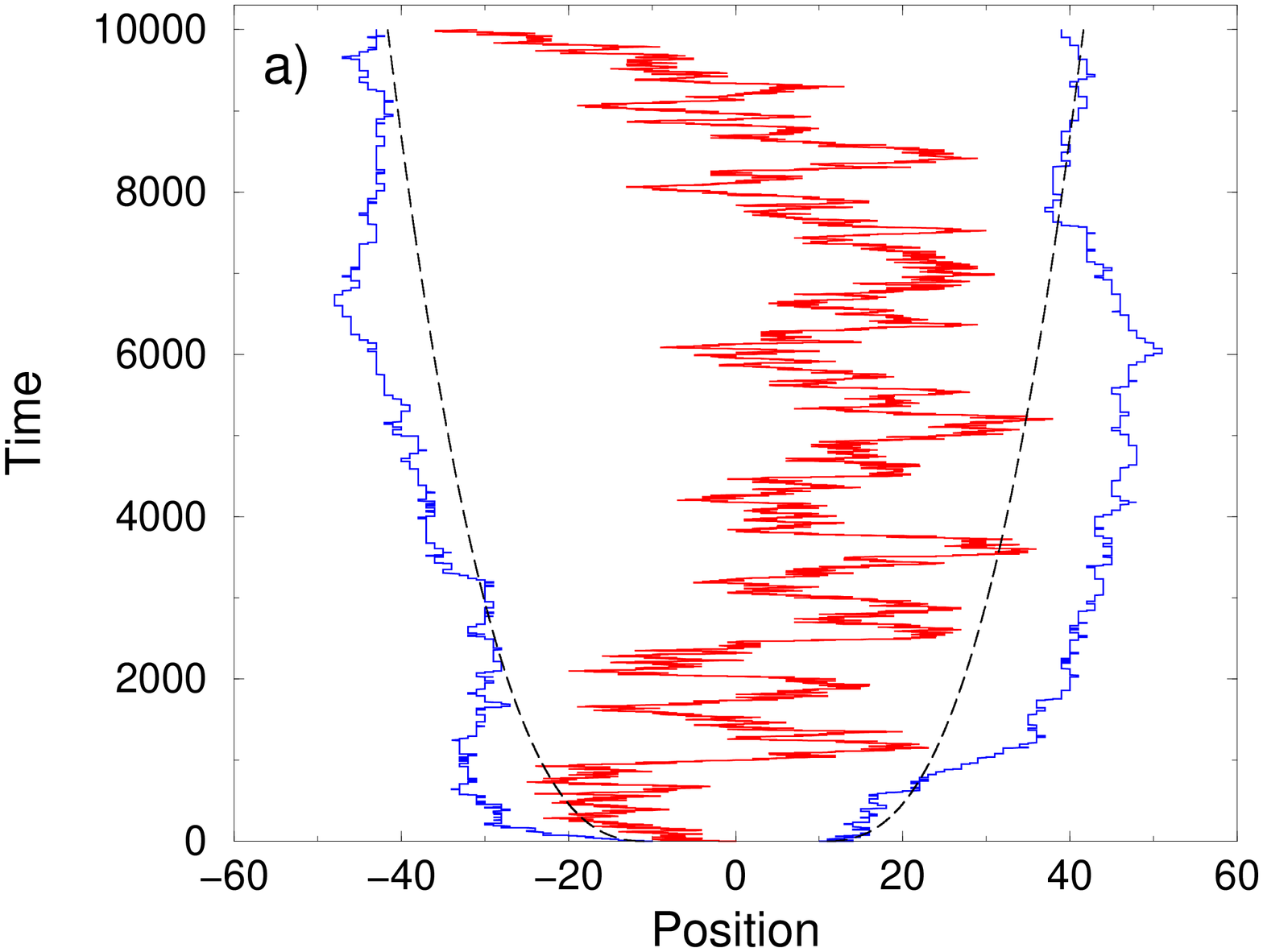}
}}
        \end{minipage}\hfill
        \vspace{1ex}
        \centering
        \begin{minipage}[b]{.48\linewidth}
                \centerline{\hbox{
\includegraphics[angle=-90,width=\linewidth]{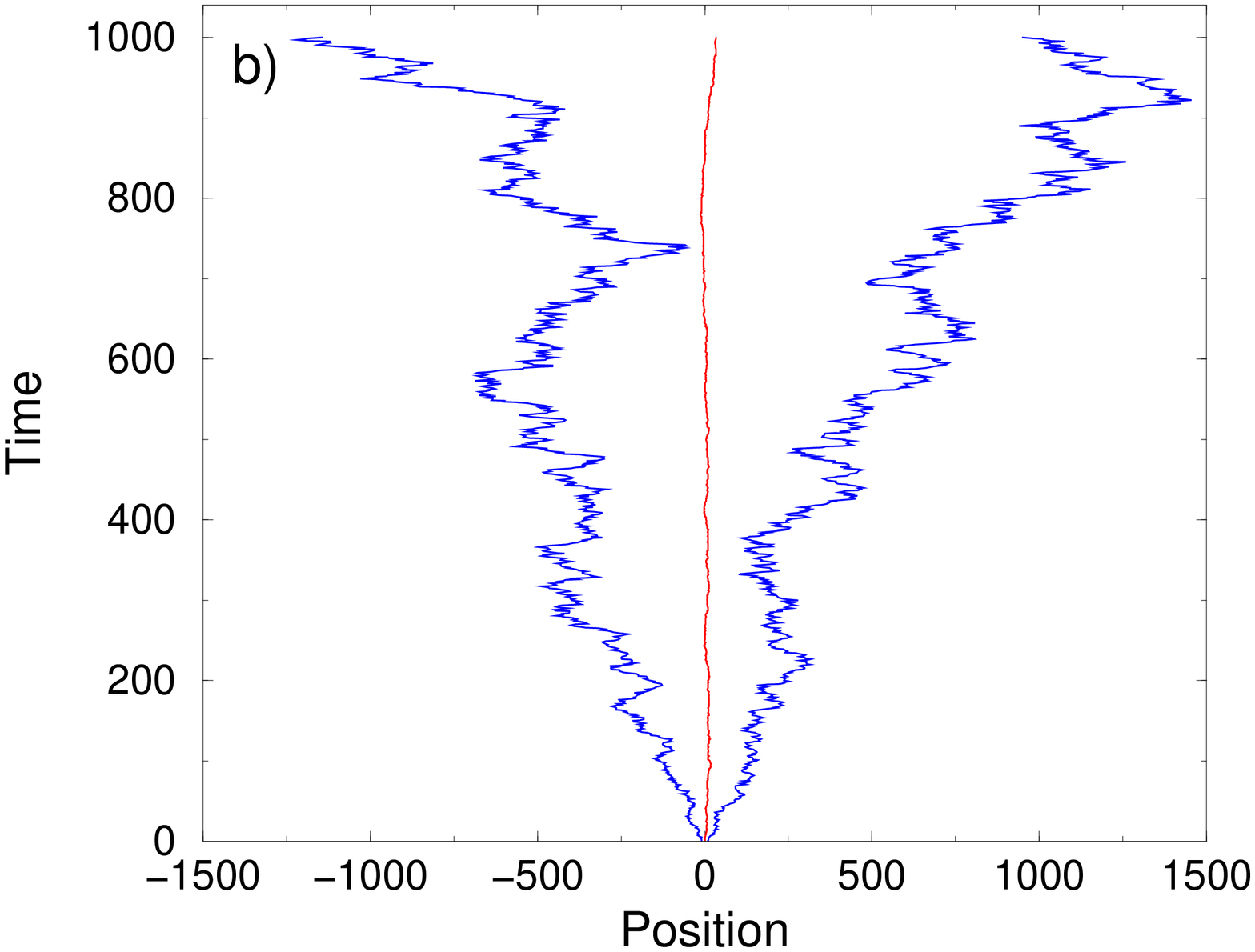}
}}
        \end{minipage}\hfill
\caption{Visualization of the three particle system when the initial 
distance $l_0=10$ for a) $\gamma = -2$ ($z=1/4$) and b) $\gamma=1$
($z=1$). The probabilities of these configurations are of order
$10^{-8}$ and $10^{-4}$, respectively. At the final time the ratio
$D_2/(D_1 t^{\gamma z})$ is about $10^2$ in (a) and $10^{-3}$ in
(b). The dashed lines in (a) show $t^\alpha$-behavior with
$\alpha=0.375$ (see Section~\ref{gammapk0} for details).
}
\label{pmainoskuva}
\end{figure*}

The three particle simulation is similar to that of the DLCA. 
Initially the distance between particles is $l_0$. A particle $i \in
\{1,2,3\}$ is selected randomly and it is moved
a distance $a$ either to the left or to the right with probability
${\cal D}_i(t)/{\cal D}_>(t)$. Here ${\cal D}_>(t) = {\rm
max}\{{\cal D}_1(t),D_2,{\cal D}_3(t)\}$ is the maximum of the diffusion
coefficients of the three particles at that time. The distance $a$ is
set to correspond the lattice constant of the DLCA simulations, \ie,
$a=1$. Irrespective of the movement of the particle time is increased
by $1/[3{\cal D}_>(t)]$ and the time-dependent diffusion coefficients
${\cal D}_1(t)$ and ${\cal D}_2(t)$ are updated to new values. This
procedure continues until a collision occurs. Figure~\ref{pmainoskuva}
shows examples of configurations that survive for a long while
for negative and positive values of the diffusion exponent. 

The faster the survival probability decays the more computation time
is used in simulating systems, which terminate at early
times. In order to sample efficiently the long living,
interesting configurations we use a cloning
method~\cite{Grassberger1,Grassberger2} for the three 
particle simulations when $\gamma < 0$: At times $t_j$ we make
$n_j$ copies of all the systems, which have survived upto this
time. Typically simulations are averaged over $5\times 10^6$
initializations and the system is copied at times $10$, $10^2$, and
$10^3$ with $2\times10^3$, $5\times10^4$, and $10^4$ copies,
respectively. This enables us to reach probabilities less than
$10^{-15}$.

\subsection{Size-Independent Diffusion ($\gamma =0$) and Crossover
Behavior} \label{gammaon0}

When the diffusion constant of a cluster does not depend on its size,
\ie\  $\gamma = 0$, an exact solution 
is possible as the collisions of
the clusters surrounding a persistent one with other clusters do
not matter~\cite{Spouge:PRL60}. 
For the same reason the mean-field approximation becomes exact and
reduces to an old problem of the survival probability of three similar
annihilating random walkers~\cite{Fisher:JSP34}. The persistence and
survival exponents attain the value $3/2$. 

This result can also be obtained from the
equation~\eqref{FPeq} which for this particular case simplifies to 
\begin{equation}
\frac{\partial{\rho}}{\partial t} = 
( 
 \frac{\partial^2{\rho}}{\partial x_{12}^2} 
 + \frac{\partial^2{\rho}}{\partial x_{23}^2}
)
- \frac{\partial^2{\rho}}{\partial x_{12} \partial x_{23}},
\end{equation}
where we have taken $D_1 = D_2 = 1/2$. 
A coordinate transformation 
$ x = (x_{12} + x_{23}), y = (x_{12} - x_{23})/\sqrt{3} $
reduces this to a diffusion equation
\begin{equation}
\frac{\partial{\rho}}{\partial t} = 
\frac{\partial^2{\rho}}{\partial x^2} 
+ \frac{\partial^2{\rho}}{\partial y^2} \label{diffeq}
\end{equation}
with the boundary condition $\rho = 0$ along lines $y= \pm
x/\sqrt{3}$. This corresponds to a two-dimensional wedge of angle
$\Theta = \pi/3$, in which the survival probability decays as
$t^{-\pi/2 \Theta} \sim t^{-3/2}$~\cite{rednerinkirja}. 

It is also interesting to know
how the asymptotic regime, where $P_{\rm surv}(t) \sim
t^{-\theta_{\rm RW}}$, is reached. 
In the case $D_{1}=D_{2}=D_{3}=D$ ($\gamma = 0$) with the initial
distances between particles being $x_{12}^0=x_{23}^0=l_0$ 
the solution including
the first correction to scaling is given
by~\cite{Derrida:PRE54,Spouge:PRL60} 
\begin{equation}
P_{\rm surv}(t) \approx
\frac{1}{4\sqrt{2\pi}}\left(\frac{l_0^2}{Dt}\right)^{3/2}
\left(1-\frac{3}{16} \frac{l_0^2}{Dt} 
\right). 
\label{correctiontozero}
\end{equation}
The correction becomes negligible for times much larger than
the crossover time 
$t_{\rm cr} = 3l_0^2/(16D)$. 
For $\gamma \neq 0$ the corrections go in powers of the ratio of the
diffusion coefficients, $D_2/D_1t^{\gamma z}$. For $\gamma >0$
this is demonstrated in Appendix~\ref{wobound} and for the
corresponding two particle problem it may be shown exactly (see
Appendix~\ref{twoparticles}). Therefore the
crossover time depends on $\gamma$ as $t_{\rm cr} \sim
r^{(2-\gamma)/|\gamma|}$, where the constant $r \approx 30$ according to 
simulations. As $t_{\rm cr}$ diverges for $|\gamma| \to
0$, we can expect that the asymptotic scaling regime can be reached in
simulations only for relatively large values of $|\gamma|$.

\subsection{Validation of the Mean-Field Theory ($\gamma > 0$)}
\label{gammask0}

We have not been able to solve equation~\eqref{FPeq}
exactly. The reason is that the absorbing boundary conditions 
together with the two time scales 
appearing in the problem make the standard methods (Laplace or Fourier
transforms; polar coordinates) unapplicable. Nor is it possible to
transform the equation 
to a diffusion equation with simple enough  boundary
conditions. However, the full solution is not needed for
the determination of the survival exponent since this is
given by the leading large time behavior when $t \to \infty$. It would only
provide us information about the crossover effects, which according to
our analysis (see Appendix~\ref{wobound}) and
the numerical simulations (see below) are rather pronounced when
$\gamma$ is close to zero.

A change of variables
$x = (x_{12} + x_{23})/\sqrt{2}, y = (x_{12} - x_{23})/\sqrt{2} $
transforms Eq.~\eqref{FPeq} to 
\begin{equation}
\frac{\partial{\rho}}{\partial t} = 
D_1t^{\gamma z}\frac{\partial^2{\rho}}{\partial x^2} 
+ (D_1t^{\gamma z} + 2D_2)\frac{\partial^2{\rho}}{\partial y^2} 
\label{diffeq2}
\end{equation}
with the boundary condition $\rho = 0$ along $y = \pm x$,
\ie, a wedge of angle $\Theta = \pi /2$. When $\gamma > 0$ the
constant term is negligible at long times ($D_1 t^{\gamma z} \gg D_2$)
and the diffusion becomes isotropic. 
This can be shown by directly solving equation~\eqref{FPeq} and
analyzing the large time behavior of the solution
(Appendix~\ref{wobound}). A change to the time scale $T_1$
[see Eq.~\eqref{T1timescal}] transforms Eq.~\eqref{diffeq2} to
the form of Eq.~\eqref{diffeq} and the survival probability $P_{\rm
surv}(t) \sim T_1^{-\pi/2\Theta} \sim 
T_1^{-1} \sim t^{-(1+\gamma z)}$. As $z=1/(2-\gamma)$ the
survival exponent $\theta_{\rm RW}(\gamma) = 2/(2-\gamma) = 2z$. 

The approximation of neglecting the constant term in Eq.~\eqref{FPeq} 
corresponds to a complete separation of the time scales, \ie, to a
situation, where the middle particle is at rest ($D_2 = 0$). Thus for 
$\gamma > 0$ one
could simply determine the survival exponent by considering two
independent random walkers with a {\it fixed} absorbing boundary in
between [compare to Fig.~\ref{pmainoskuva}~(b)]. In other
words, the motion of the ``slow'' particle becomes 
asymptotically irrelevant. This can be exactly shown for the the
corresponding two particle problem (Appendix~\ref{twoparticles}).

\begin{figure*}
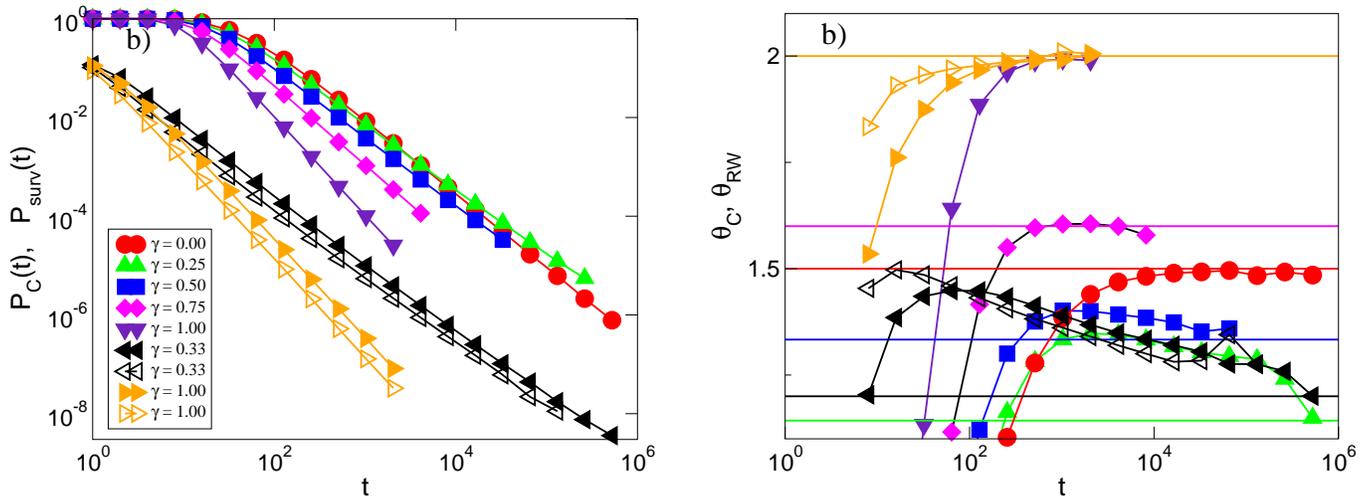

        \centering
        \begin{minipage}[b]{.48\linewidth}
                \centerline{\hbox{
\includegraphics[width=\linewidth]{jul2posg_insetu.eps}
}}
        \end{minipage}\hfill
        \vspace{1ex}
        \centering
        \begin{minipage}[b]{.48\linewidth}
                \centerline{\hbox{
\includegraphics[width=\linewidth]{jul2posg_inset2.eps}
}}
        \end{minipage}\hfill
\caption{a) Comparison between the survival (filled symbols) and
 persistence (open symbols) probabilities.
b) The corresponding local exponents. The
horizontal lines correspond to the analytic values given 
by $\theta = 2/(2-\gamma)$. The data for RW survival are averaged over
variable number of realizations ranging from $10^9$ for $\gamma = 0$
to $2\times10^7$ for $\gamma = 0.5$. The DLCA simulations are averaged
over 50000 simulations on a system of size 55555. The initial distance between
particles is $10$ (upper curves in Fig. a) or $2$ (lower curves). 
}
\label{surv1fig}
\end{figure*}

Figure~\ref{surv1fig} compares the survival and persistence
probabilities. The initial distances between particles in the random
walk simulations are set to be the same as in the DLCA.
The probabilities decay algebraically at large times and the only
difference in the decay is between the amplitudes. This is to be
expected as the transient effects of the growth of the average cluster
size are not taken into account in the random walk picture. 

The inset shows local exponents, \ie\ logarithmic derivatives of the
probabilities, which converge to the value obtained 
from the Fokker-Planck equation, $\theta_{\rm RW} = 2/(2-\gamma)$ for $\gamma
>0$ and $\theta_{\rm RW} = 3/2$ for $\gamma=0$. 
The asymptotic regime is reached only for $\gamma=0$ and $\gamma
\gtrsim 0.5$. In the latter region the 
local exponents saturate, when the ratio of the diffusion 
coefficients is of about $30$. For example, for $\gamma = 0.25$ this
would corresponds to $t_{\rm cr} \approx 2 \times 10^{10}$ 
which is beyond the time reached in simulations. 

Note that the persistence exponent is discontinuous
and nonmonotonic at $\gamma = 0$, \ie, $3/2 = \theta_C(0) >
\theta_C(0^+) =  1$. This seems first counterintuitive since making
some of the clusters to diffuse faster helps others to survive
longer! On the other hand, as time elapses a persistent cluster
becomes slower as compared to an average one. In this way it
eventually adopts the optimal
strategy~\cite{Redner:AJP67} by becoming stationary.

\subsection{Fluctuation dominated persistence ($\gamma < 0$)} \label{gammapk0}

For $\gamma < 0$ the diffusion of the clusters surrounding a
persistent one slows down.  
Consider the random walk picture and proceeding similarly
as for $\gamma >0$ above. 
Fixing now particles $1$ and $3$ would lead to an interval of fixed length 
and hence to an exponentially decaying
survival probability. However, simulations show
that the survival decays stretched exponentially in time,
$P_{\rm surv}(t) \sim \exp (-C_{\rm RW} t^{\beta_{\rm RW}})$. 
Furthermore, as will be shown below, although
the surrounding particles become slower, their motion can not be
neglected even at the long time limit. This is a collective effect
and in clear contrast to the exactly solvable two particle case,
where the fast particle eventually dominates the survival
(Appendix~\ref{twoparticles}).  

\begin{figure}
\centering
\includegraphics[width=\linewidth
]{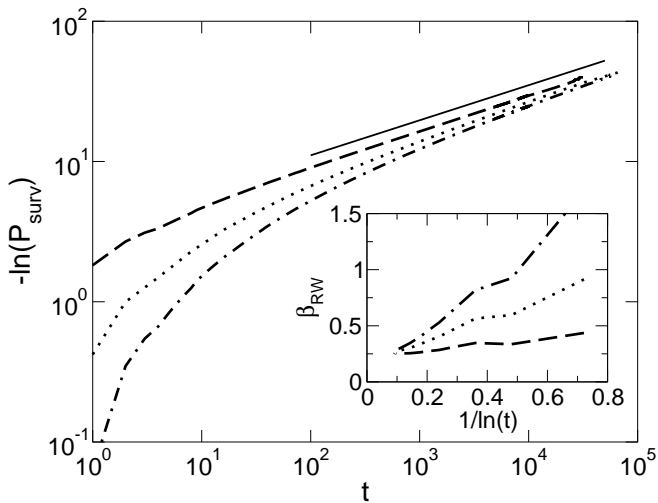}
\caption{Survival probabilities for 
$\gamma = -2$ with $l_0 = 2$~(dashed), $3$~(dotted), and
$4$~(dot-dashed). The solid line is a guide to eye with a slope
$\beta_{\rm RW} = 0.25$. The inset shows how the local stretching
exponents converge to the same value independent of the initial
distance $l_0$. 
}
\label{survnegfig}
\end{figure}

In figure~\ref{survnegfig}
we plot $-\ln (P_{\rm surv}(t))$ vs. $t$ on a log-log 
scale so that a stretched exponential decay corresponds to a straight
line with a slope $\beta_{\rm RW}$. The final slope is independent of
the initial distance between particles, and thus the
stretching exponent is universal.

\begin{figure}
\centering
\includegraphics[width=\linewidth
]{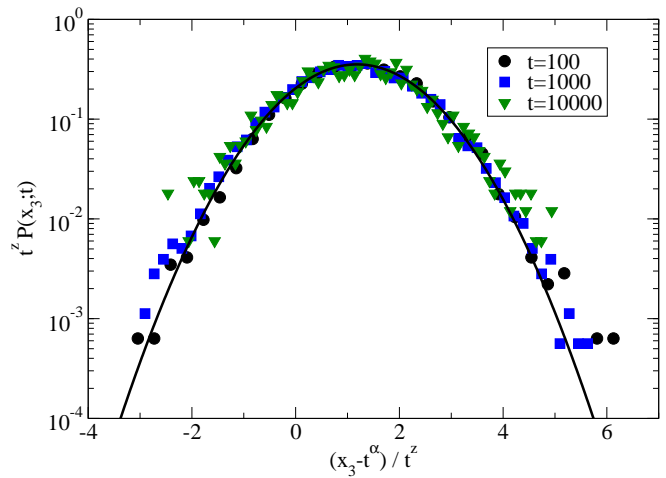}
\caption{The scaling plot of the location distribution of the
rightmost particle in the random walk simulations for $\gamma=-2$.
The values of the scaling exponents are $z=1/4$ and $\alpha = 3/8$.
The solid line shows a Gaussian fit to the data. 
}
\label{neg_gamma_scal_for3}
\end{figure}

Figure~\ref{neg_gamma_scal_for3} shows the location
distribution $p(x_3;t)$ of the particle 3 (the one for the particle 1
would be the same). It scales as 
\begin{equation}
p(x_3;t) = t^{-z} g\left( \frac{x-bt^\alpha}{t^z} \right)
\label{x3scaleq} 
\end{equation}
implying that although the distribution widens as $t^z$, the
expectation value of the distance from the origin  
grows as $bt^\alpha$ with a non-trivial exponent $z < \alpha < 1/2$
[see Fig.~\ref{pmainoskuva}~(a)]. 
The scaling is similar to the reaction front in the originally
separated reaction-diffusion 
system $A+B \to C$, where the reaction zone becomes sharp at
late times, \ie\ $z<\alpha$~\cite{Galfi:PRA88,Koza:PHYSA240}. It is
striking that the scaling function $g(y)$ is within the numerical
accuracy a simple Gaussian. 

The consequence of Eq.~\eqref{x3scaleq} is that the average
distance between the particles 1 and 3 grows [see
Fig.~\ref{pmainoskuva}~(a)].  
If it would grow {\it deterministically\/} as $t^{\alpha}$, 
with $\alpha < 1/2$, the survival probability would decay
asymptotically stretched exponentially with the exponent 
$\beta_{\rm det} = 1-2\alpha$~\cite{Krapivsky:AJP64}. 
For example, for $\gamma = -2.0$ the numerics gives a rough estimate
$\alpha \approx 0.36$ and $1-2\alpha \approx 0.28$, which is in
reasonable agreement with the numerically obtained stretching esponent
$\beta_{\rm RW} \approx 0.25$ (see inset of Fig.~\ref{survnegfig}).

To understand the origin of the
new length scale $t^\alpha$ the next 
logical step is to try to take the length fluctuations 
of the interval into account. We make this using 
a Lifshitz tail approach~\cite{rednerinkirja}. It is based on the
assumption that 
the main contribution to the survival is provided by extreme
configurations, where the particles surrounding the surviving one
have diffused far apart from each other. We write the survival
probability as 
\begin{equation}
P_{\rm surv}(t) \approx \int_0^\infty {\rm d}l\ P(l;t) Q(t|l),
\label{Lifeq}
\end{equation}
where $P(l;t)$ is the probability distribution of the interval lengths
$l=x_3-x_1$ around a surviving particle at time $t$ and $Q(t|l) \sim
l^{-1}\exp(-\pi^2 D t/l^2)$ is the survival probability of a particle in an
interval of length $l$~\cite{rednerinkirja,Weiss:book}. 
In order to make progress, we need to know the large $l$
behavior of $P(l;t)$. It scales similarly as $p(x_3;t)$ 
\begin{equation}
P(l;t) = t^{-z} G\left( \frac{l-2bt^\alpha}{t^z} \right),
\label{Pltscaleq} 
\end{equation}
where the large $y$ tail of $G(y)$ is Gaussian as 
the position distributions of particles 1 and 3 are Gaussian.
Although it is irrelevant in what follows, the small $y$ part of $G(y)$
decays faster than the large $y$ tail due to the restriction 
$x_3 > x_1$.

Denote the variance of the Gaussian tail of $G(y)$ by
$\sigma^2$. Then Eq.~\eqref{Lifeq} gives
\begin{equation*}
P_{\rm surv}(t) \sim 
t^{z-\alpha} \int_{0}^\infty {\rm d}l \exp { \left( 
-\frac{(l-2bt^{\alpha})^2}{2\sigma^2 t^{2z}} -\frac{\pi^2 D t}{l^2}
\right)}.
\end{equation*}
When $t \to \infty$ the integrand becomes sharply peaked and may be
evaluated using the saddle point method. This gives $\alpha = (2z+1)/4$
and  
\begin{equation}
P_{\rm surv}(t) \sim t^{(6z-1)/4}e^{-C t^{(1-2z)/2}}.
\end{equation}
Inserting the value of $\alpha$ coming from the Lifshitz argument
to the result of an algebraically expanding interval,
$\beta_{\rm det} = 1-2\alpha$, leads to the same streching exponent
$\beta = (1-2z)/2$. These two
results coincide, as a consequence of the peculiar scaling
[Eq.~\eqref{Pltscaleq}] and that the tail of the interval length
distribution decays as $G(y) \sim \exp(-y^2)$. We emphasize that the
fluctuations of the surrounding, slow particles determine the
stretching exponent and that it is purely a coincidence that 
the Lifshitz tail argument gives the same result as the use of the
average value.

The stretching exponent $\beta_{\rm RW} = (1-2z)/2$ has an obvious
interpretation. There are two length scales in the problem. The first one 
is related to the random walkers with time-dependent diffusion
coefficients, $L_1 \sim t^z$, and the other to the surviving particle,
$L_2 \sim t^{1/2}$. The argument of the exponential decay is simply
the ratio of these two scales in the problem, $P_{\rm surv}(t) \sim
\exp(-L_2/L_1)$. Although reasonable, the calculation above shows
the delicacy of the survival: the distance between the particles 1 and
3 involves a third, non-trivial length scale, $L_3 \sim t^\alpha$,
with $\alpha = (2z+1)/4$. The above
considerations can also be made by resorting to an argument which
considers the two
characteristic time scales $T_1 \sim t^{1+\gamma z}$ and $T_2 \sim
t$. It is easy to see, that the ratios between the scales
obey a diffusive like scaling relation $L_2/L_1 \sim \sqrt{T_2/T_1}$
such that any quantity involving the ratio of length scales may 
be given in terms of the ratio of the time scales and vice versa.

\begin{figure}
\centering
\includegraphics[width=\linewidth
]{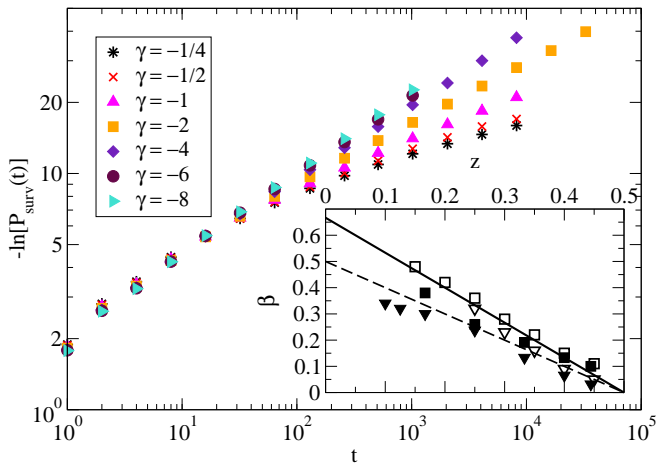}
\caption{The survival probability for $\gamma < 0$. The inset shows
the bounds for the stretching exponents for the survival (filled
symbols) and persistence (open symbols). For details see text. The
dashed [solid] line is given by $(1-2z)/2$ [$2(1-2z)/3$].
}
\label{survpersfiggammalt0}
\end{figure}

In figure~\ref{survpersfiggammalt0} the survival probabilities are
plotted for $\gamma < 0$ (for a similar figure for the persistence see
figure~3 in~\cite{Hellen:EPL}). 
In spite of being able to simulate rather small probabilities 
the asymptotic regime is not reached in the simulations. 
Similar problems with a slow convergence to the asymptotic
value have been encountered in other reaction-diffusion
systems~\cite{Mehra:preprints,Bray:preprint} and they might be
overcome by a more 
efficient use of the cloning method~\cite{Grassberger1,Grassberger2}. 
The inset of Fig.~\ref{survpersfiggammalt0} shows bounds for the
stretching exponents as a function of the dynamic exponent
$z=1/(2-\gamma)$. The upper bounds are obtained by
fitting a line to the three or four largest time points and measuring the
slope. To obtain the lower bound, we considered the change of the
local slope and extrapolated to $1/t \to 0$, when it was
possible. This method neglects the saturation of the local exponent after
a finite crossover time and therefore gives a lower bound. For comparison, 
the corresponding bounds for persistence are also shown in the inset. 
There is a clear difference between the two. 
The numerics is consistent with the prediction $\beta_{\rm RW} =
(1-2z)/2$, and for the persistence the data suggest an expression 
$\beta_C = 2(1-2z)/3$.

\begin{figure}
\centering
\includegraphics[width=\linewidth
]{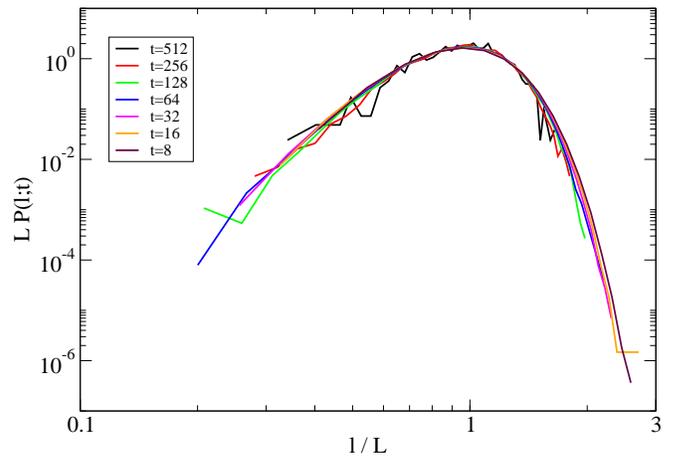}
\caption{The scaling of the distance distribution between the clusters
surrounding a persistent one in DLCA for $\gamma = -0.8$. }
\label{distdistforDLCA}
\end{figure}

The difference between the mean-field model and the DLCA is further
elucidated in figure~\ref{distdistforDLCA}.
It shows that in the DLCA the distance distribution between the
clusters surrounding a persistent one scales 
similar to that of the cluster size distribution
\begin{equation}
P(l;t) = L^{-1} h \left( \frac{l}{L} \right). \label{DLCAPltscal}
\end{equation}
Hence, the  distribution widens at the same rate as the average
distance $L(t) \sim t^z$ grows in contrast to the RW case. 
For large $x$ the scaling function $h(x) \sim \exp(-bx)$  and the
Lifshitz tail argument leads to an estimate $\beta_{\rm L} =
(1-2z)/3$, which disagrees with the numerics.  

The inconsistency is not surprising since in the DLCA there are
fluctuations coming from the statistical nature of collisions, which
are not taken into account in the Lifshitz approach. More precisely,
the diffusion constants of the neighbors of persistent clusters have
some unknown distribution. Furthermore, the diffusion constant also 
correlates
with the distance from the persistent cluster. These facts together
with the fact that the stretching exponent is determined by
the fluctuations makes an analytical estimation of the persistence
for $\gamma < 0$ hard.

\section{Implications for the Cluster Size Distribution} \label{poly}

We now turn to the relation of the persistence
to the cluster size distribution. We concentrate first on the case 
$\gamma > 0$,
when the cluster size distribution has a power-law tail at small
cluster sizes. The dynamical scaling together with the definition
of the dynamical and polydispersity exponents, $z$ and $\tau$,
respectively, were discussed in the Introduction 
[Eq.~\eqref{nstscaleq}]. The scaling theory further states that all
the cluster number densities decay in a similar manner at large 
times, \ie, $n_s(t)/n_1(t) \to b_s$ as $t \to
\infty$~\cite{vanDongen:JPA18}, where $b_s$ is a 
constant.  Here the exponent of interest 
is the universal decay exponent, $w$, which describes the decrease
$n_s(t) \sim t^{-w}$.

Using $S(t) \sim t^z$ together with $f(x) \sim x^{-\tau}$ as $x \to 0$
in equation~\eqref{nstscaleq} gives $n_s(t) \sim t^{-(2-\tau)z}
s^{-\tau}$ so that the three exponents defined above are related by
the scaling relation
$w = (2-\tau)z$~\cite{Vicsek:PRL52}. 
Therefore the full characterization of the dynamic scaling requires
the knowledge of only two of the exponents.
However, even on
the mean-field level of Smoluchowski's rate equation theory the only
readily calculable exponent for DLCA is the dynamic exponent
$z$. The difficulty with, for example, the polydispersity
exponent $\tau$ arises from the fact that to calculate it 
requires the knowledge of the whole scaling
function~\cite{vanDongen:PRL54}.  
Next we argue how knowing the persistence exponent $\theta_C$
helps to overcome this problem.

Let us start from the trivial size independent case, $\gamma = 0$, for
which an exact solution of the cluster size distribution $n_s(t)$ is
possible~\cite{Spouge:PRL60}. The actual form of this
distribution is not important for our purposes. The
point is that the decay exponent $w=3/2$ for any short-range correlated
initial distribution $n_s(0)$. Also the cluster persistence exponent
is universal~\cite{omapers}. 
Hence, by noticing that for a
monodisperse initial condition, $n_s(0) = \delta_{1,s}$, the
persistence probability is simply $n_1(t)$, we obtain the persistence
exponent $\theta_C(0) = w(0) = 3/2$.

The exponents $\theta_C$ and $\tau$
should be related also for $\gamma \neq 0$, since
the persistent clusters are those ones, which have not
aggregated with other ones. Asymptotically, the number of these
clusters will be presented by the part $s \ll S(t)$ of the cluster
size distribution, which in turn is characterized by the exponent
$\tau$. 
Thus the same identification $\theta_C = w$ can be made also for $0
< \gamma < 2$ and we are led to the scaling relation 
\begin{equation}
\theta_C = (2-\tau)z. \label{scalrelC}
\end{equation}
The same relation is valid in a different context of the scaling of
intervals between persistent regions in the reaction-diffusion model
$A+A \to \emptyset$~\cite{Manoj:JPA33}. 
Here $\theta_C = 2z$ and Eq.~\eqref{scalrelC} gives $\tau(\gamma) = 0$. 
This is interesting in two respects. First,
the polydispersity exponent is discontinuous as $\gamma \to 0$ since
$\tau(0) = -1  \neq  0 = \tau(0^+)$. Although quite uncommon,
such an outcome is possible also on the mean-field level of the rate equation
theory~\cite{vanDongen:PRA32}. 
It is more surprising, that the polydispersity exponent is a constant,
independent of the value of $\gamma$. It indicates that for any
$\gamma >0$ the physics of small clusters is dictated by the fact
that they are essentially immobile compared to the larger (average-sized)
ones in the system.

\begin{table}
\caption{
\label{exptable}
Exponents measured from the numerical
data. For $\gamma= 0.40$ the asymptotic regime is not reached in
simulations (except for $z$) and only upper bounds are shown. 
}  
\begin{ruledtabular}
\begin{tabular}{cccc}
$\gamma$ & $z$ & $\theta_C$ & $\tau$ \\
\hline
$0.00$ & $0.500 \pm 0.001$ & $1.50 \pm 0.02$ & $1.00 \pm 0.02$ \\
$0.40$ & $0.625 \pm 0.001$ & $<1.35$ & $<0.10$ \\
$0.57$ & $0.699 \pm 0.002$ & $1.43 \pm 0.05$ & $0.02 \pm 0.05$ \\
$1.00$ & $1.00 \pm 0.01$   & $2.00 \pm 0.02$ & $0.00 \pm 0.02$ \\
\end{tabular}
\end{ruledtabular}
\end{table}

Simulations confirm the constant value of $\tau$ although again the
crossover effects make the analysis intractable near $\gamma =
0$~\cite{Hellen:EPL}. 
The numerically estimated values for the exponents are presented in
table~\ref{exptable}. The scaling relation~\eqref{scalrelC}
is obeyed within the error bars.

For $\gamma < 0$ the scaling function of the cluster size distribution
behaves as $f(x) \sim \exp (-A x^{-|\mu|})$ when $x \to 0$. Using
a similar reasoning 
as for $\gamma \ge 0$ leads now to the relation $\beta_C  = |\mu|z$.  
Together with the result $\beta_C = 2(1-2z)/3$ this suggests that
$\mu(\gamma) = 2 \gamma /3$. Direct measurement of the exponent $\mu$
is hard as one would need to compute the scaling function $f(x)$ for
$x \lesssim 0.1$ to see the asymptotic behavior. However, even rough 
numerics shows that $\mu(-2) > -1.75$, which is
larger than the mean-field value predicted by the Smoluchowski's rate
equation theory, $\mu = \gamma$. Hence, the spatial fluctuations help
clusters to survive longer.

\section{Conclusions} \label{conclusion}

We have investigated the probability of a cluster to remain
unaggregated in one-dimensional DLCA. The diffusivity of clusters
is taken to vary with size as $D(s) \sim s^\gamma$, which extends
the results known for $\gamma = 0$ to the more relevant case of
size dependent diffusion. 

The first main result is that the persistence probability decays as 
\begin{equation}
P_{\rm surv}(t)  \sim
\begin{cases}
\exp(-C t^{\beta_C})  &, \gamma < 0 \\
t^{-3/2}  &, \gamma = 0 \\
t^{-2/(2-\gamma)}  &, \gamma > 0. \label{varthetagamma2}
\end{cases}
\end{equation}
The stretching exponent fits well to the expression $\beta_C =
2(1-2z)/3$ where the dynamic exponent is given by $z=1/(2-\gamma)$. 
Equation~\eqref{varthetagamma2} shows that one can not use the exactly
solvable size independent aggregation as a starting point
for a perturbative analysis of the size dependent case. 
The second main result is that the decay of persistence is related to
the dynamic exponent $z$ through the scaling relations 
\begin{eqnarray*}
\theta_C &=& (2-\tau)z \\
\beta_C  &=& |\mu|z,
\end{eqnarray*}
where the exponents $\tau$ and $\mu$ characterize the small size tail
of the cluster size distribution. Hence, by solving for the
persistence one determines the behavior 
of the cluster size distribution. 
For $\gamma \ge 0$ the scaling relation and Eq.~\eqref{varthetagamma2}
lead to a discontinuity of the polydispersity exponent: $\tau(0) = -1$
but for $0 < \gamma < 2$ the distribution is flat and $\tau = 0$.  

The persistence probability for $\gamma \ge 0$ is obtained from a
mean-field analysis for three annihilating random walkers. It explains
the discontinuous and non-monotonic 
behavior of the persistence exponent, \ie, why $3/2 = \theta_C(0) >
\theta_C(0^+) = 1$. 
This is since for $\gamma >0$ a persistent cluster
eventually adopts the optimal strategy~\cite{Redner:AJP67} by becoming
more and more stationary as time goes on. This interpretation is further
supported by the fact that the probability of an originally empty site
to be never occupied by a cluster decays algebraically with the same
exponent as the cluster persistence~\cite{omapers}. The major consequence
of the discontinuity is the divergence of the 
crossover time to the asymptotic behavior when $\gamma \to
0^+$. This also plagues the scaling of the cluster size
distribution since these two are interconnected. 

The mean-field random walk analysis, which can be analyzed in
the asymptotic limit when $\gamma \ge 0$, becomes intractable
for $\gamma < 0$. We have thus resorted to numerical studies. These
reveal that while the RW picture adequately describes the
persistence for $\gamma \ge 0$, it is inadequate for $\gamma < 0$. The
reason is that there the persistence is affected by the fluctuations
in the motion of the slowly moving particles around the persistent
one. These are taken into account approximately in the
mean-field theory, which results only to a qualitative understanding
of the persistence. For $\gamma > 0$ the approximation is practicable
as the fluctuations of the slow particle/clusters become asymptotically
irrelevant. For $\gamma < 0$ they are significant as the persistence
decays much faster than a power law. As an interesting consequence,
the mean-field theory is applicable 
when the cluster size distribution is broad around the mean
($\gamma \ge 0$: $f(x) \sim x^{-\tau}$, $x\to 0$) but not when it is narrow
($\gamma <0$; $f(x)$ decays faster than any power for $x\to
0$). 

The difference between the mean-field random walk model and the DLCA is
demonstrated by the scaling of the distribution measuring the distance
between the particles (clusters) enclosing a surviving (persistent)
one [see Eqs.~\eqref{Pltscaleq} and \eqref{DLCAPltscal}]. The main
difference is that in the theory the average distance grows faster
than the distribution widens whereas in the DLCA these both take place
at the same rate. This implies the existence of a new, non-trivial
length scale $\sim t^{\alpha}$ in the RW-problem. 
A Lifshitz tail argument suggest an expression $\alpha =
(2z+1)/4$. This leads to $\beta_{\rm RW} = (1-2z)/2$, which agrees
with the simulations. Hence, the argument of the exponential decay is
the ratio of the two natural length scales, $t^{1/2}$ and $t^z$, of
the problem. An intriguing detail of the random walk model is that 
according to the numerics the position distribution of the neighbor of
the surviving particle scales as $p(x;t) = t^{-z} g \left(
[x-bt^\alpha]/t^z  \right)$ with a purely Gaussian scaling function
$g(y)$. It would be worthwhile to try to show this analytically and
also solve Eq.~\eqref{FPeq} with appropriate boundary conditions. This
would require new analytic tools to handle time-dependent absorbing
boundary value problems as the traditional image method can not be
applied. We believe this to be an unsolved mathematical problem
waiting for solution. 

The present study investigates cluster persistence in diffusion--limited
cluster-cluster aggregation. It would be interesting to consider
the behavior of unaggregated clusters in other models, too. 
Furthermore, we have concentrated only on
the one-dimensional case. It is natural to ask what can be done in
higher dimensions.  There a similar simple random walk analysis is
hardly possible. On the other 
hand the long crossover effects near $\gamma = 0$ presumably persist
and make simulation studies hard. Nevertheless, we believe that
the general structure of the problem remains and conclude with
the conjecture that also in
higher dimensions the behavior of the cluster size distribution is
determined by the solution of the cluster persistence problem. 

\acknowledgments

E.K.O.H thanks F.~Leyvraz and S.~Redner for proposing the Lifshitz
tail argument for the survival when $\gamma < 0$.

\appendix

\section{Asymptotic analysis of survival for $\gamma > 0$}
\label{wobound} 

The Fourier transform of Eq.~\eqref{FPeq} reads
\begin{equation}
\frac{\partial{\hat{\rho}}}{\partial t} = 
-(D_2+D_1t^{\gamma z}) (k_x^2 + k_y^2) \hat{\rho}
+ 2 D_2 k_x k_y \hat{\rho},
\label{FtFPeq} 
\end{equation}
where we have for notational simplicity used variables $x$ and $y$
instead of $x_{12}$ and $x_{23}$, respectively. The hat denotes the
Fourier transform and $k_x$ and $k_y$ are the associated Fourier
variables of $x$ and $y$. 
The solution of Eq.~\eqref{FtFPeq} fulfilling the initial condition
$\rho_f(x,y;0) = \delta(x-x_0) \delta(y-y_0)$ is 
\begin{equation*}
\hat{\rho}_f(k_x,k_y;t) = e^{i k_x x_0 + i k_y y_0
-D_2 t (k_x-k_y)^2-{\cal D}(t)t(k_x^2+k_y^2)},
\end{equation*}
where ${\cal D}(t) = D_1 t^{\gamma z} / (\gamma z+1)$.
The subscript $f$ refers to the solution without absorbing boundaries.  
The inverse transform reduces to calculating Gaussian
integrals with the result
\begin{widetext}
\begin{equation} 
\rho_f(x,y;t|x_0,y_0) = 
\frac{1}{4\pi t 
\sqrt{{\cal D}(t) \left[ 2 D_2 + {\cal D}(t) \right]}}
\exp \left( -\frac{
	\left[ D_2 + {\cal D}(t) \right] 
	\left[ (x-x_0)^2 + (y-y_0)^2 \right] + 
2D_2(x-x_0)(y-y_0)}{4 t {\cal D}(t) \left[ D_2 + {\cal D}(t) \right]}
\right).  \label{rhosol}
\end{equation}
\end{widetext}
At the long time limit this reduces to a Gaussian 
\begin{equation*}
\rho_f^{\rm as}(x,y;t|x_0,y_0) = \frac{1}{4\pi {\cal D}(t)t}
\exp \left( -\frac{ (x-x_0)^2 + (y-y_0)^2} {4 {\cal D}(t) t} \right),
\end{equation*}
which is nothing but the solution of Eq.~\eqref{FPeq} for $D_2 =
0$. This validates the approximation made in section~\ref{gammask0}.

Since the solution~\eqref{rhosol} is not symmetric in reflection
with respect to the 
$x$- and $y$-axis, the method of images frequently used in problems
including absorbing boundaries can not be applied to construct the 
solution which would be zero along the axes. 
To obtain an estimate for the survival probability as a series
expansion in powers of $t$, we 
neglect the cross-term $2D_2(x-x_0)(y-y_0)$ in the exponential of
Eq.~\eqref{rhosol} and denote the resulting radially symmetric part by
$\rho^{\rm S}_f$. The term omitted is of the same order in $t$ as the
term $D_2[ (x-x_0)^2 + (y-y_0)^2]$ and would hence
contribute only on the prefactors in the expansion. 
Now the image method gives the solution 
obeying $\rho = 0$ along $x=0$ and
$y=0$ for $x \ge 0$ and $y \ge 0$:
\begin{eqnarray}
\rho(x,y;t|x_0,y_0) &\approx& 
\rho_f^{\rm S}(x,y;t|x_0,y_0) \nonumber \\
&-& \rho_f^{\rm S}(x,y;t|-x_0,y_0) \nonumber \\ 
&-& \rho_f^{\rm S}(x,y;t|x_0,-y_0) \nonumber \\ 
&+& \rho_f^{\rm S}(x,y;t|-x_0,-y_0). \nonumber
\end{eqnarray}
Integrating this over the first quadrant \{$x \ge 0, y \ge 0$\} yields
\begin{equation}
P_{\rm surv}(t) \approx \frac{2zx_{0}y_{0}}{\pi D_{1}t^{2z}}
\left[1-2z{\cal R}+6z^2{\cal R}^2+ {\cal O}({\cal R}^3)\right],
\label{fullsolexpansion3}
\end{equation}
where ${\cal R}=D_2/(D_1t^{\gamma z})$ denotes the ratio of the diffusion
coefficients.  The asymptotic behavior sets in for ${\cal R} \ll 1$,
which indicates the divergence of the crossover time to the asymptotic
behavior when $\gamma \to 0$.

\section{Two particle survival} \label{twoparticles} 

Consider the survival of two particles, which annihilate at contact
but otherwise evolve according to 
\begin{equation}
\begin{cases}
\dot{x}_1(t) &=  \xi_1(t) \\
\dot{x}_2(t) &=  \xi_2(t), 
\end{cases} \nonumber \label{Langfor2part}
\end{equation}
where $\langle \xi_i(t) \rangle
= 0$ and $\langle \xi_i(t) \xi_j(t') \rangle =  2 {\cal D}_i(t)
\delta_{ij} \delta(t-t')$. 
Let the diffusion coefficients of the particles to be ${\cal D}_1(t)
= D_1 t^{\gamma z}$ and ${\cal D}_2(t) = D_2$, where $z =
1/(2-\gamma)$.  
The distance $y(t)=x_2(t)-x_1(t)$ between the particles obeys the
Langevin equation
\begin{equation}
\dot{y}(t) =  \sqrt{D_1t^{\gamma z}+D_2}\Gamma(t), \nonumber
\label{distsfor2part} 
\end{equation}
where $\langle \Gamma(t) \rangle = 0$ and $\langle \Gamma(t) \Gamma(t')
\rangle =  2 \delta(t-t')$. This is of the standard
form~\cite{Risken:book} and can directly be transformed to a
Fokker-Planck equation 
\begin{equation}
\frac{\partial{\rho(y;t)}}{\partial t} = (D_2+D_1t^{\gamma z})
 \frac{\partial^2{\rho(y;t)}}{\partial y^2},
\label{FP2peq} 
\end{equation}
where $\rho(y;t)$ is the probability density of finding the
two particles at distance $y$ at time $t$. 

The solution of Eq.~\eqref{FP2peq} fulfilling the boundary and initial
conditions $\rho(0;t) = 0$ and $\rho(y;0) = \delta(y-y_0)$ is readily found
to be
\begin{equation}
\rho(y;\Ttt(t)) = \frac{1}{\sqrt{4\pi \Ttt}}\left[ e^{-(y-y_0)^2/4\Ttt} -
e^{-(y+y_0)^2/4\Ttt} \right], \nonumber
\end{equation}
where $\Ttt (t) = D_2 t + D_1 t^{1+\gamma z}/(1+\gamma z)$.
The survival probability 
\begin{equation}
P_{\rm surv}(t) = \int_0^\infty {\rm d}y\ \rho(y;t) =
{\rm erf}\left(\frac{y_0}{\sqrt{4 \Ttt}}\right) 
\nonumber
\end{equation}
whose asymptotic behavior at large $t$ is given by
\begin{equation}
P_{\rm surv}(t) \sim 
\begin{cases}
y_0 (\pi D_2 t)^{-1/2} \left[ 1 - {\cal R}/(4z) + \ldots \right] &,
\gamma < 0\\ 
y_0 [\pi (D_1+D_2) t]^{-1/2} \left[ 1 - {\cal O}(t)\right]&, \gamma = 0 \\
y_0 \left[ \pi D_1 t^{2z}/2z\right]^{-1/2} 
\left[ 1 - z/{\cal R} + \ldots \right] &, \gamma > 0,
\end{cases} \nonumber
\end{equation}
where ${\cal R}=D_2/(D_1t^{\gamma z})$ illustrating again the
divergence of the crossover time when $|\gamma| \to 0$. 
The survival exponent $\theta_{\rm RW} = {\rm max}\{1/2,z\}$, \ie, it is
given by the dynamics of the faster particle. The
interpretation of the result for $\gamma \neq 0$ is simple: eventually
the time scales separate and the slower particle becomes
stationary.

\end{document}